\documentclass{article}
\usepackage{amsmath,color}
\usepackage[english]{babel}
\usepackage{latexsym}
\usepackage{amssymb}
\usepackage{graphicx}

\newcommand{\tmop}[1]{\ensuremath{\operatorname{#1}}}
\newcommand{\tmtextbf}[1]{{\bfseries{#1}}}
\newcommand{\tmtextit}[1]{{\itshape{#1}}}
\definecolor{grey}{rgb}{0.75,0.75,0.75}
\definecolor{orange}{rgb}{1.0,0.5,0.5}
\definecolor{brown}{rgb}{0.5,0.25,0.0}
\definecolor{pink}{rgb}{1.0,0.5,0.5}

\begin{document}

\title{Non Abelian Fields in Very Special Relativity}
\author{Jorge Alfaro, \cr
Facultad de F\'\i sica, Pontificia Universidad
Cat\'olica de Chile,\cr
 Casilla 306, Santiago 22, Chile\cr 
 jalfaro@puc.cl \cr
 and \cr
Victor O. Rivelles,\cr
Instituto de F\'\i sica, Universidade de S\~ao Paulo, \cr
C. Postal 66318,  S\~ao Paulo, SP, Brazil\cr
rivelles@fma.if.usp.br }

\maketitle
\begin{abstract}
We study non-Abelian fields in the context of very special relativity (VSR).
For this we define the covariant derivative and the gauge field gauge
transformations, both of them involving a fixed null vector $n_{\mu}$, related
to the VSR breaking of the Lorentz group to the Hom(2) or Sim(2) subgroups. As
in the Abelian case the gauge field becomes massive. Moreover we show that the
VSR gauge transformations form a closed algebra. We then write actions
coupling the gauge field to various matter fields (bosonic and fermionic). We
mention how we can use the spontaneous symmetry breaking mechanism to give a
flavor dependent VSR mass to the gauge bosons. Finally, we quantize the model
using the BRST formalism to fix the gauge. The model is renormalizable and
unitary  and for non abelian groups, asymptotically free.
\end{abstract}

\section{Introduction}

Special relativity (SR) is valid at the largest energies available today \cite{auger}.
However the possible violation of the underlying Lorentz symmetry presents us with new experimental and theoretical challenges.
In particular, its violation has been considered as a possible evidence for Planck scale Physics, as some  theories of quantum gravity predicts it \cite{alfaro1}.
Experiments and astrophysical observations are used to set
stringent bounds upon the parameters describing these violations. Broadly speaking, three  basic scenarios have been explored: (1) Non-dynamical
tensor fields are introduced to determine preferred directions that break the
Lorentz symmetry. Some instances of this are the Myers-Pospelov model {\cite{MP}} together with QED in a constant axial vector background {\cite{aacgs}}. (2) A second scenario is to assume  spontaneous symmetry
breaking (SSB) of the Lorentz symmetry as  in the standard model extension of {\cite{KOSTELECKY0}},{\cite{REV}}, where such non-dynamical tensor fields are
assumed to arise from vacuum expectation values of some basic fields belonging
to a more fundamental theory. (3) A third possibility has been introduced in {\cite{cohen1}}. 

There it is proposed that the laws of nature are not
invariant under the whole Lorentz group (with 6 parameters) but instead are
invariant under subgroups of the Lorentz group which still preserves the basic elements of SR like the constancy of the velocity of light. It was named very special relativity (VSR). In general space isotropy and CP are violated but if CP is incorporated as a symmetry then the whole Lorentz group is recovered. The most
interesting of these subgroups are Sim(2) (with 4 parameters) and Hom(2), (with 3
parameters). These subgroups do not have invariant tensor fields besides the
ones that are invariant under the whole Lorentz group, implying that the
dispersion relations, time delay and all classical tests of SR are valid for
these subgroups too. New effects would be generated by parity violating and non-local terms.
VSR admits the generation of a neutrino mass without lepton number violation
nor sterile neutrinos {\cite{cohen2}}. The implications of this novel
mechanisms could be tested at non-relativistic neutrino energies such as the
end point of the electron spectrum in beta decay. 
VSR has been generalized to include supersymmetry \cite{Cohen:2006sc}, curved spaces \cite{Gibbons:2007iu}, noncommutativity \cite{SheikhJabbari:2008nc}, cosmological constant {\cite{enrique}}, dark matter \cite{Ahluwalia:2010zn}, cosmology \cite{Chang:2013xwa} and Abelian gauge fields {\cite{chun}}.
Some doubts about the  phenomenology of Hom(2) VSR has been presented in 
{\cite{Das}}, arguing that it  is unable to explain Thomas precession, and further extended in {\cite{us}}.  

In this paper we want to consider non-Abelian gauge fields in the context of
VSR. We review the results of the Abelian case {\cite{chun}} in section 2. In section 3, we define the covariant derivative, the gauge transformations
and the action for non-Abelian fields. 
Section 4 is then devoted to SSB of the VSR gauge symmetry, section 5 introduces the BRST gauge fixing, in section 6 we discuss renormalizability and unitarity and section 7 contains the conclusions and open problems.

\section{Abelian Gauge Fields}

In this section, we follow closely {\cite{chun}}. 
Let us consider a  gauge field $A_{\mu}$ in VSR with gauge transformation
\begin{equation} \label{1}
  \delta A_{\mu} = \tilde{\partial}_{\mu} \Lambda, 
\end{equation}
where the wiggle operator is defined by:
\begin{equation}
  \tilde{\partial}_{\mu} = \partial_{\mu} - \frac{1}{2} \frac{m^2}{n\cdot
  \partial} n_{\mu},
\end{equation}
and $n\cdot\partial = n^\mu \partial_\mu$. The constant vector $n^\mu$ is a null vector and transforms multiplicatively under a VSR transformation so that any term containing ratios involving $n^\mu$ are invariant. In order to have the usual mass dimension for $\tilde{\partial}_\mu$ a constant $m$ has to introduced and sets the scale of VSR effects. 

Consider now a charged scalar field $\phi$ to be coupled to the gauge field and with gauge transformation
\begin{align}
	 \delta \phi = i \Lambda \phi. 
\end{align}
It can be shown that the operator $D_{\mu}$ {\cite{chun}}:
\begin{equation}
  D_{\mu} \phi = \partial_{\mu} \phi - i A_{\mu} \phi - \frac{i}{2} m^2
  n_{\mu} ( \frac{1}{(n\cdot  \partial)^2} n\cdot A) \phi,
\end{equation}
satisfies the fundamental property of transforming as $\phi$ does under gauge
transformations:
\begin{eqnarray*}
  \delta (D_{\mu} \phi) = i \Lambda D_{\mu} \phi, &  & 
\end{eqnarray*}
We will call $D_\mu \phi$ the covariant derivative of $\phi$.
Associated to this covariant derivative, we define the wiggle covariant derivative of the field $\phi$ by 
\begin{equation}
  \tilde{D}_{\mu} \phi = D_{\mu} \phi - \frac{1}{2}  \frac{m^2}{n\cdot D} n_{\mu}
  \phi. \label{wcd}
\end{equation}
It reduces to $\tilde{\partial}_{\mu} \phi$ for $A_{\mu} = 0$. We will see below that using the wiggle covariant derivative in the action
gives to $\phi$ a \ mass{\footnote{The action of a scalar field using the
wiggle operator (2) gives the field a Lorentz invariant mass $m$. This can be
seen integrating by parts to get the action $- \int d^d x \phi (x)
\tilde{\partial}^2 \phi (x) = - \int d^d x \phi (x) (\Box - m^2) \phi (x)$}}.


The field strength related to $D_\mu$ can be computed as 
\begin{eqnarray*}
  {}[D_{\mu}, D_{\nu}] \phi & = - i F_{\mu \nu} \phi,  & 
\end{eqnarray*}
and it is given by 
\begin{equation}
  F_{\mu \nu} = \partial_{\mu} A_{\nu} - \partial_{\nu} A_{\mu} + \frac{1}{2}
  m^2 n_{\nu} ( \frac{1}{(n\cdot  \partial)^2} \partial_{\mu} (n\cdot A)) - \frac{1}{2}
  m^2 n_{\mu} ( \frac{1}{(n\cdot  \partial)^2} \partial_{\nu} (n\cdot A)). 
  \label{AbelianF}
 \end{equation}
We call  $F_{\mu\nu}$ as the $A_\mu$
 field strength.
It does not coincide with 
\begin{equation}
 \tilde{\partial}_{\mu} A_{\nu} - \tilde{\partial}_{\nu} A_{\mu}, 
\end{equation}
which is also gauge invariant and will be used below to describe massive gauge fields.
However, the difference between them must be gauge invariant 
\begin{eqnarray*}
  \tmop{dF}_{\mu \nu} = F_{\mu \nu} - ( \tilde{\partial}_{\mu} A_{\nu} -
  \tilde{\partial}_{\nu} A_{\mu})  =\frac{1}{2} m^2\left( n_{\nu} \frac{1}{(n\cdot \partial)^2}
  n_{\alpha} F_{\mu \alpha} - n_{\mu} \frac{1}{(n \cdot\partial)^2} n_{\alpha}
  F_{\nu \alpha}\right). 
\end{eqnarray*}
We then define the $A_{\mu}$ wiggle field strength by
\begin{equation}
  \tilde{F}_{\mu \nu} = F_{\mu \nu} - \frac{1}{2} m^2\left( n_{\nu} \frac{1}{(n\cdot  \partial)^2}
  n_{\alpha} F_{\mu \alpha} - n_{\mu} \frac{1}{(n\cdot  \partial)^2} n_{\alpha}
  F_{\nu \alpha} \right), \label{abelianfs}
\end{equation}
where  by construction:
\begin{equation}
  \tilde{F}_{\mu \nu} = \tilde{\partial}_{\mu} A_{\nu} -
  \tilde{\partial}_{\nu} A_{\mu}. \label{vsrF}
\end{equation}

Using the wiggle covariant derivative (\ref{wcd}) we can write a gauge invariant
action for the charged scalar coupled to the Abelian gauge field. Since $\tilde{F}_{\mu\nu}$ is gauge invariant the action can have, besides the usual $\tilde{F}^2$ term, contributions involving the square of  $n_\mu \tilde{F}^{\mu\nu}$, so that the most general gauge invariant action quadratic in the gauge field is 
\begin{equation} 
  S = \int d^n x \left( - \frac{1}{4} \tilde{F}_{\mu \nu} \tilde{F}^{\mu \nu}
  + \frac{g}{2} \frac{1}{n\cdot  \partial} n_{\alpha} \tilde{F}^{\alpha \mu}
  \frac{1}{n\cdot  \partial} n_{\alpha} \tilde{F}^{\alpha}_{\mu} + |
  \tilde{D}_{\mu} \phi |^2 \right),  \label{Abelianaction}
\end{equation}
where $g$ is a constant. 
That the action (\ref{Abelianaction}) describes a massive gauge field can mostly easily be seen 
in the simpler case where $g=0$ and disregarding the coupling to the scalar field. 
The free Abelian field equation of motion derived from $S$ is then  
\begin{equation}
\tilde{\partial}^{\mu} \tilde F_{\mu \nu} = 0.
\end{equation}
Choosing as gauge condition a VSR type Lorentz gauge $\tilde{\partial}^{\mu} A_{\mu} = 0$, we get
   \begin{eqnarray*}
  \tilde{\partial}^2 A_{\nu} =  ( \Box - m^2) A_{\nu} = 0. & &
\end{eqnarray*}
i.e. $A_\mu$ has mass $m$. A similar argument shows that the scalar field also has mass $m$. 

Similarly for fermions coupled to an Abelian gauge field we have the gauge
invariant lagrangian
\begin{equation}
 \mathcal{L}= \bar{\psi} \gamma^{\mu} i \tilde{D}_{\mu} \psi,
 \end{equation}
and again it is possible to show that the fermionic field has mass $m$. 

In order to handle the non-local terms we use the definition 
\begin{equation}
 \frac{1}{n\cdot\partial} = \int_0^{\infty} d a \,\,  e^{- a n\cdot\partial}. 
 \end{equation}

Notice that replacing the wiggle by the raw definitions still preserve the symmetry of the action but now describes massless particles instead of VSR massive particles.

\section{Non Abelian Gauge Fields}

This is the most important section of the paper. We obtain the generalization
of the covariant derivative and the gauge transformations in the presence of a
non-Abelian gauge field in VSR. 


We consider a scalar field transforming under a non-Abelian gauge transformation
with infinitesimal parameter $\Lambda$
\begin{equation}
     \delta \phi = i \Lambda \phi. 
   \end{equation}
As before, we define the covariant derivative by
\begin{equation}
  D_{\mu} \phi = \partial_{\mu} \phi - i A_{\mu} \phi - \frac{i}{2} m^2
  n_{\mu} ( \frac{1}{(n\cdot  \partial)^2} n\cdot A) \phi. \label{cdna}
\end{equation}
To find out the non-Abelian gauge transformation of the gauge field we write 
\begin{equation}
 \delta A_{\mu} = \partial_{\mu} \Lambda - i [A_{\mu}, \Lambda] + f_{\mu}, 
 \end{equation}
and $f_\mu$ is determined by imposing the proper transformation property for the
covariant derivative,
\begin{eqnarray*}
  \delta (D_{\mu} \phi) = i \Lambda D_{\mu} \phi. &  & 
\end{eqnarray*}
We then get
\begin{eqnarray*}
  f_{\mu} &=& \frac{i}{2} m^2 n_{\mu} \Lambda ( \frac{1}{(n\cdot  \partial)^2} n\cdot A) -
  \frac{1}{2} m^2 n_{\mu} ( \frac{1}{(n\cdot  \partial)} \Lambda) \nonumber \\
  &+& \frac{i}{2} m^2
  n_{\mu} ( \frac{1}{(n\cdot  \partial)^2} n\cdot  [A, \Lambda]) - \frac{i}{2} m^2
  n_{\mu} ( \frac{1}{(n\cdot  \partial)^2} n\cdot A) \Lambda. 
\end{eqnarray*}
Then the gauge transformation for the gauge field is
\begin{eqnarray}
  \delta_{\Lambda} A_{\mu} = \partial_{\mu} \Lambda - i [A_{\mu}, \Lambda] &+& 
  \frac{i}{2} m^2 n_{\mu} \left[ \Lambda, ( \frac{1}{(n\cdot  \partial)^2} n\cdot A)
  \right] - \frac{1}{2} m^2 n_{\mu} ( \frac{1}{(n\cdot  \partial)} \Lambda) \nonumber \\ &+&
  \frac{i}{2} m^2 n_{\mu} ( \frac{1}{(n\cdot  \partial)^2} n\cdot  [A, \Lambda])  
\end{eqnarray}
For an Abelian gauge field $A_{\mu}$ we get (\ref{1}). 
We have also checked the closure of the algebra
\begin{eqnarray*}
  {}[\delta_{\Lambda_1}, \delta_{\Lambda_2}] A_{\mu} = - i \delta_{[\Lambda_1,
  \Lambda_2]} A_{\mu} &  & 
\end{eqnarray*}

We also define the wiggle covariant derivative of the field $\phi$ by:
\begin{equation}
  \tilde{D}_{\mu} \phi = D_{\mu} \phi - \frac{1}{2} \frac{m^2}{n\cdot D} n_{\mu}.
  \phi \label{wcdna}
\end{equation}
It reduces to $\tilde{\partial}_{\mu} \phi$ for $A_{\mu} = 0$.


The commutator of two covariant derivatives defines $F_{\mu \nu}$, the $A_\mu$ field strength,
\begin{equation}
  [D_{\mu}, D_{\nu}] \phi = - i F_{\mu \nu} \phi, \label{fs1}
\end{equation}
so we get 
\begin{eqnarray}
  &F_{\mu \nu} &= A_{\nu, \mu} - A_{\mu, \nu} - i [ A_{\mu}, A_{\nu}] +
  \frac{1}{2} m^2 n_{\nu} ( \frac{1}{(n\cdot  \partial)^2} n\cdot A_{, \mu}) -\nonumber \\ &&
  \frac{1}{2} m^2 n_{\mu} ( \frac{1}{(n\cdot  \partial)^2} n\cdot A_{, \nu}) -
   \frac{i}{2} m^2 \left[ ( \frac{1}{(n\cdot  \partial)^2} n\cdot A), ( n_{\mu} A_{\nu} -
  n_{\nu} A_{\mu}) \right]. \label{F}
\end{eqnarray}
It is hermitian if $A_\mu$ is hermitian and it coincides with (\ref{AbelianF}) for Abelian $A_{\mu}$.
\begin{eqnarray*}
  {}[D'_{\mu}, D'_{\nu}] \phi' = U [D_{\mu}, D_{\nu}] \phi = U (- iF_{\mu
  \nu}) \phi = (- iF'_{\mu \nu}) U \phi, &  & U = e^{i \Lambda}
\end{eqnarray*}
we find
\begin{equation}
  F'_{\mu \nu} = UF_{\mu \nu} U^{- 1} .
\end{equation}

The non-Abelian generalization of (\ref{abelianfs}) is
\begin{eqnarray}
  \tilde{F}_{\mu \nu} = F_{\mu \nu} - \frac{1}{2} m^2\left( n_{\nu} \frac{1}{(n\cdot D)^2} (
  n_{\alpha} F_{\mu \alpha}) - n_{\mu} \frac{1}{(n\cdot D)^2} ( n_{\alpha} F_{\nu
  \alpha}) \right), &  &  \label{nafs}
\end{eqnarray}
where $D_{\mu}$ is the covariant derivative acting on fields that
transform in the adjoint representation (see (\ref{28}) below).
Using the wiggle covariant derivative (\ref{wcdna}) we can write a gauge invariant
action for the scalar field coupled to a  non-Abelian gauge field
\begin{equation}
  S = \int d^n x \left[ - \frac{1}{4} \tmop{tr} \left( \tilde{F}_{\mu \nu}
  \tilde{F}^{\mu \nu} + \frac{g}{2} \frac{1}{n\cdot D} n_{\alpha} \tilde{F}^{\alpha
  \mu} \frac{1}{n\cdot D} n_\beta \tilde{F}^\beta_\mu \right) + |
  \tilde{D}_{\mu} \phi |^2 \right]. \label{NonAbelianaction}
\end{equation}
Similarly, for fermions coupled to the non-Abelian gauge field we have the
gauge invariant Lagrangian
\begin{equation}
  \mathcal{L}= \bar{\psi} \gamma^{\mu} i \tilde{D}_{\mu} \psi.
\end{equation}
We again use the definition
\begin{equation}
 \frac{1}{n\cdot D} = \int_0^{\infty} d a e^{- a n\cdot D},
\end{equation}
so that
\begin{eqnarray}
  \tilde{F}_{\mu \nu} = \tilde{\partial}_{\mu} A_{\nu} -
  \tilde{\partial}_{\nu} A_{\mu} + {\cal O} ( A^2). &  &  \label{linearF}
\end{eqnarray}
Notice that $\tilde{F}_{\mu \nu}$ is the right
field strength  to describe massive bosons (and not $F_{\mu \nu}$).
That is (\ref{NonAbelianaction}) describes massive gauge bosons (when
$g = 0$) because the term quadratic in $A_{\mu}$ in (\ref{NonAbelianaction}) is the
Abelian one (see (\ref{Abelianaction}) with $g = 0$) and as shown earlier it describes massive gauge bosons. 


When the scalar field is in the adjoint representation we have 
\begin{eqnarray}
 \delta \phi = i [ \Lambda, \phi]. 
\end{eqnarray}
The covariant derivative now takes the form 
\begin{equation}\label{28}
  D_{\mu} \phi = \partial_{\mu} \phi - i [ A_{\mu}, \phi] - \frac{i}{2} m^2
  n_{\mu} \left[ ( \frac{1}{(n\cdot  \partial)^2} n\cdot A), \phi \right],
  \end{equation}
 while the non-Abelian gauge field transforms as
 \begin{eqnarray} \label{29}
  \delta A_{\mu} = \partial_{\mu} \Lambda - i [A_{\mu}, \Lambda] &+& \frac{i}{2}
  m^2 n_{\mu} \left[ \Lambda, ( \frac{1}{(n\cdot  \partial)^2} n\cdot A) \right] -
  \frac{1}{2} m^2 n_{\mu} ( \frac{1}{(n\cdot  \partial)} \Lambda) \nonumber \\
  &+& \frac{i}{2} m^2
  n_{\mu} ( \frac{1}{(n\cdot  \partial)^2} n\cdot  [A, \Lambda]).
\end{eqnarray}
From these we get the proper transformation for the covariant derivative
\begin{equation}
 \delta ( D_{\mu} \phi) = i [ \Lambda, D_{\mu} \phi]. 
 \end{equation}

\section{Spontaneous Symmetry Breaking}

In the context of VSR the non-Abelian gauge bosons all have mass $m$. 
In order to give different masses to the gauge bosons we need to break the
non-Abelian gauge symmetry. For this we consider a scalar field in the adjoint
representation of the gauge group coupled to the gauge field
\begin{equation}
  S = \int d^n x \left[ - \frac{1}{4} \tmop{Tr} ( \bar{F}_{\mu \nu}
  \bar{F}^{\mu \nu}) + \tmop{Tr} ( D_{\mu} \phi)^2 + V ( \phi) \right],
  \label{ssbaction}
\end{equation}
where we defined
\begin{eqnarray}
  \bar{F}_{\mu \nu} = F_{\mu \nu} - \frac{1}{2} \mu^2 \phi^2 \left( n_{\nu}
  \frac{1}{(n\cdot D)^2} ( n_{\alpha} F_{\mu \alpha}) - n_{\mu} \frac{1}{(n\cdot D)^2} (
  n_{\alpha} F_{\nu \alpha}) \right), &  &  \label{ssbnafs}
\end{eqnarray}
with $\mu$ being a dimensionless parameter.
If $\phi$ gets a vacuum expectation value $v$, then the gauge field will get a
VSR mass matrix
\[ M = \mu v \]
in addition to the usual mass matrix coming from:
\[ \tmop{Tr} ( [ A_{\mu}, v])^2 \]
In the Standard Model $\mu$ must be  tiny  since there is no
evidence of violations of SR there.

\section{BRST transformations}

In this section we derive the BRST transformation for the non-abelian gauge
field in VSR. We follow {\cite{kugo}} closely introducing the ghosts $C, \bar{C}$ and $B$. Taking into account the gauge transformation (\ref{29}) we find the nilpotent BRST transformations 
%
\begin{eqnarray}
  \delta A_{\mu} &=& \tilde{\partial}_{\mu} C - i [A_{\mu}, C] +  \frac{i}{2}
  m^2 n_{\mu} \left[ C, ( \frac{1}{(n \cdot \partial)^2} n \cdot A) \right] \nonumber \\
  & &+ \frac{i}{2} m^2 n_{\mu}  (\frac{1}{(n \cdot \partial)^2} n \cdot [A,
  C]),  \nonumber \\
  \delta C &=& i C^2,  \nonumber\\
  \delta \bar{C} &=& i B,   \nonumber\\
  \delta B &=& 0.
\end{eqnarray}


To fix the gauge, we add to the action (\ref{NonAbelianaction})(without the scalar field) the following term
\begin{eqnarray*}
  S_{\tmop{gh} + \tmop{FP}} &=& - i \delta \tmop{tr} \left\{ \left(
  \tilde{\partial}_{\mu} A^{\mu} + \frac{1}{2} \alpha B \right) \bar{C}
  \right\}  \nonumber \\
  &=&- i \tmop{tr} \left\{ - \bar{C} (\Box - m^2) C + i \bar{C} \partial_{\mu}
  [A_{\mu}, C] - \frac{1}{2} i m^2  \bar{C} \left( \frac{1}{(n \cdot \partial)}
  [n.A, C] \right) \right. \nonumber \\
	&-&\left. \frac{i}{2} m^2 \bar{C}  \left( (n. \partial) \left[ C, (
  \frac{1}{(n \cdot \partial)^2} n \cdot A) \right] \right) + i \left(
  \tilde{\partial}_{\mu} A^{\mu} + \frac{1}{2} \alpha B \right) B \right\},
\end{eqnarray*}
which generates the kinetic and interacting terms for the ghosts. 
Integrating over $B$ we get
\begin{eqnarray*}
  S_{\tmop{gh} + \tmop{FP}} &=& - i \tmop{tr} \left\{ - \bar{C} (\Box - m^2)
  C + i \bar{C} \partial_{\mu} [A_{\mu}, C] - \frac{1}{2} i m^2  \bar{C}
  \left(\frac{1}{(n \cdot \partial)} [n.A, C]\right) \right. \nonumber \\ 
	&-& \left. \frac{i}{2} m^2 \bar{C}  \left(
  (n. \partial) \left[ C, ( \frac{1}{(n \cdot \partial)^2} n \cdot A) \right]
  \right) - \frac{i}{2 \alpha} (\tilde{\partial}_{\mu} A^{\mu})^2 \right\}.
\end{eqnarray*}
Notice that the ghost propagator is 
\[ \vartriangle_{\tmop{gh}} (p) = - \frac{1}{p^2 + m^2}, \],
so that the ghost has mass $m$.

We also see that the $\bar{C} - A - C$ has corrections proportional to $m^2$.
These corrections match the behavior of the corresponding $\bar{C} - A - C$
in the standard Lorentz invariant Yang- Mills theory for large momentum.

The VSR Yang-Mills field propagator in the $\alpha$ gauge is given by 
\[ D^{\mu \nu} = \frac{1}{P^2} \left( \eta^{\mu \nu} - \frac{}{}
   \frac{1}{P^2} \frac{\alpha - 1}{2\alpha - 1} P^{\mu}
   P^{\nu} \right), \]
where $P_\mu = p_\mu + \frac{1}{2} m^2 n_\mu ( \frac{1}{(n\cdot p)}$. 
We see that for large $p$ it reduces to 
\[ D^{\mu \nu} \sim \frac{1}{p^2} \left( \eta^{\mu \nu} - \frac{}{}
   \frac{1}{p^2} \frac{\alpha - 1}{2\alpha - 1} p^{\mu}
   p^{\nu} \right), \]
and in Lorentz gauge we have 
\[ D^{\mu \nu} = \frac{\eta^{\mu \nu}}{p^2 + m^2}, \]
as expected. 

\section{Renormalization and Unitarity}

We see from the Feynman rules that propagators and vertices have the same
large momentum behaviour as in Lorentz invariant Yang- Mills theories.
Therefore the model is renormalizable.

For the same reason, non-abelian VSR Yang- Mills theories are asymptotically
free. To see this let us consider the diagrams that determines the
renormalization group $\beta$ function in the background field
method {\cite{abbott}}:

\begin{center}

\includegraphics{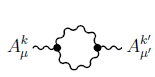}

\includegraphics{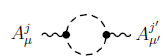}
\end{center}

\noindent The wiggle line represents the VSR Yang-Mills propagator and the discontinuous line represents the VSR ghost propagator. We see that the loop momentum integral of these graphs have the same large
momentum behavior as in the Lorentz invariant Yang-Mills model. So the
residue of the pole in dimensional regularization is the same. It follows that
both models, the VSR and the Lorentz invariant Yang-Mills, have the same
$\beta$ function and so both models are asymptotically free. 
The poles and the residues of the propagators of the particles in VSR are
the same as for Lorentz invariant theories so that unitarity is satisfied.

\section{Conclusions}

In this paper we have studied non-Abelian fields in VSR. To do this we have to
define a covariant derivative and a modified gauge transformation. We have
checked that the VSR covariant derivative commutes with the gauge symmetry.
Moreover the non-Abelian VSR gauge transformations of the gauge field form a
closed algebra. Having these, we can easily built actions for matter fields
coupled to the VSR gauge fields. One important point to notice is that the VSR
gauge fields are massive, although with a common mass. Since in nature gauge
fields may have different masses, as the Standard Model shows, we implement
the spontaneous symmetry breaking in VSR with non-Abelian gauge symmetry. In
this way the gauge fields can get the usual mass coming from spontaneous
symmetry breaking plus a flavour dependent VSR mass.

The BRST formalism was used to fix the gauge and obtain the propagators and
vertices of pure VSR Yang-Mills theory. In these models the ghosts have the
same mass of the Yang- Mills fields. Vertices and propagators match the behavior of the usual Yang-Mills vertices and
propagators for large momenta. Therefore the model is renormalizable and
asymptotically free. Also unitarity is preserved.

\section*{Acknowledgments}

The work of J.A. was partially supported by Fondecyt \# 1110378 and Anillo ACT 1102. He also wants to thank the Instituto de F\'\i sica, USP and the IFT/SAIFR for 
its kind hospitality during his visits to S\~ao Paulo. The work of V.O.R. is supported by CNPq grant 304116/2010-6 and FAPESP grant 2008/05343-5. He also wants to thank Facultad de Fisica, PUC Chile for its kind hospitality during his visits to Santiago.

\end{document}